\documentclass[lettersize,journal]{IEEEtran}
\usepackage{amsmath,amsfonts}
\usepackage[beginLComment=\#~,endLComment=~, beginComment=\#~,endComment=~,commentColor=blue]{algpseudocodex}
\usepackage{algorithm}
\usepackage{array}
\usepackage[caption=false,font=normalsize,labelfont=sf,textfont=sf]{subfig}
\usepackage{textcomp}
\usepackage{stfloats}
\usepackage{url}
\usepackage{verbatim}
\usepackage{graphicx}
\usepackage{siunitx}
\usepackage{bm}
\usepackage{multicol}
\usepackage{multirow}
\usepackage{cite}
\begin{document}

\title{A Computational Harmonic Detection Algorithm to Detect EM Side Channel Leakage from IoT Devices}
\title{A Computational Harmonic Detection Algorithm to Detect Electromagnetic Side Channel Leakage}
\title{A Computational Harmonic Detection Algorithm to Detect Unintentional Electromagnetic Leakage}
\title{A Computational Harmonic Detection Algorithm to Detect Data Leakage through EM Emanation}
\author{Md Faizul Bari,~\IEEEmembership{Student Member,~IEEE}, Meghna Roy Chowdhury,~\IEEEmembership{Student Member,~IEEE}, Shreyas Sen,~\IEEEmembership{Senior Member,~IEEE}
\thanks{The authors are with the Elmore Family School of Electrical and Computer Engineering, Purdue University, West Lafayette, Indiana 47907, USA (e-mail: mbari@purdue.edu; mroycho@purdue.edu; shreyas@purdue.edu)}
\thanks{Manuscript received October 9, 2024. This work was supported by the Office of the Director of
National Intelligence (ODNI), Intelligence Advanced Research Projects Activity (IARPA), via contract: 2021-21062400006. (Corresponding author: Md Faizul Bari)}}

\markboth{Journal of \LaTeX\ Class Files,~Vol.~x, No.~x, Oct~2024}%
{Bari \MakeLowercase{\textit{et al.}}: A Computational Harmonic Detection Algorithm to Detect Data Leakage through EM Emanation}

\IEEEpubid{0000--0000/00\$00.00~\copyright~2024 IEEE}

\maketitle

\begin{abstract}
Unintended electromagnetic emissions, called EM emanations, can be exploited to recover sensitive information, posing security risks. Metal shielding, used by defense organizations to prevent data leakage, is costly and impractical for widespread use. This issue is particularly significant for IoT devices due to their sheer volume and varied deployment environments. Therefore, there is a research need for an automated detection method to monitor facilities and address data leakage promptly. To resolve this challenge, in the preliminary version of this work \cite{date_hdmi}, a CNN-based detection method was proposed using HDMI cable emanations that provided \bm{${\sim}95\%$} accuracy up to \bm{$22.5\text{ m}$} but had limitations due to training data. In this extended version, we augment the initial study by collecting and characterizing emanation data from IoT devices, everyday electronics, and cables. We propose a harmonic-based emanation detection method by developing a computational harmonic detection algorithm. The proposed method addresses the limitations of the CNN-based method and provides \bm{${\sim}100\%$} accuracy not only for HDMI emanation (compared to \bm{${\sim}95\%$} in the earlier CNN method) but also for all other tested devices and cables. Finally, it has also been tested in different environments to prove its efficacy in practical scenarios.
\end{abstract}

\begin{IEEEkeywords}
emanation, unintended radiated emission, side channel, electromagnetic compatibility, CNN, harmonic detection, Arduino, Zigbee, EMSEC, URE
\end{IEEEkeywords}

\section{Introduction}
\begin{figure}
\centering
\includegraphics[width=3.4in]{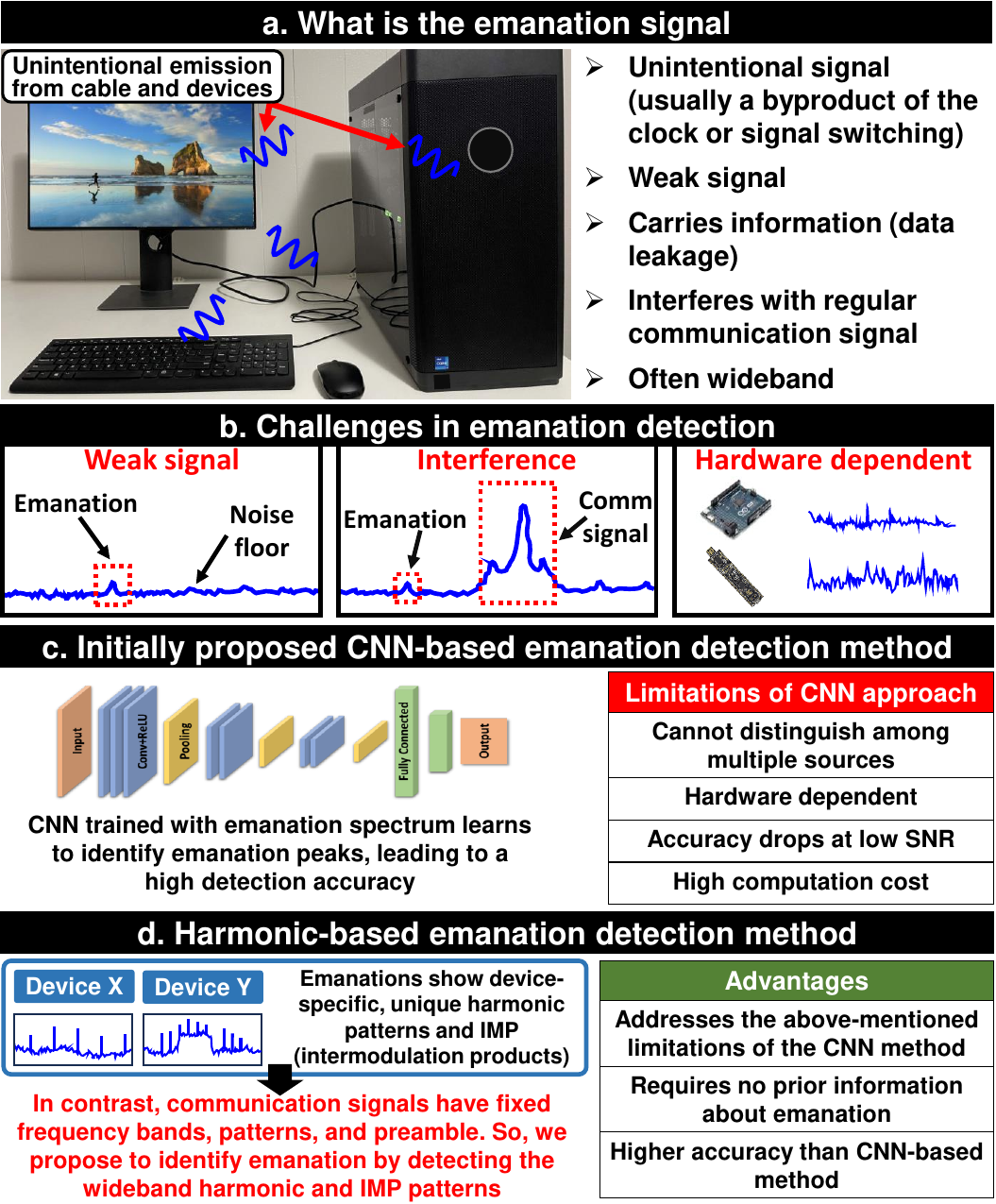}
\caption{(a) Emanation is unintentional emission from electronic devices and cables. It is a byproduct of signal switching. (b) The challenges in emanation detection stem from the fact that these are weak signals as they weren't designed to be transmitted. They are often overcast by strong interference from communication signals. (c) A CNN, trained with the processed power spectrum, can distinguish the emanation peaks from other signals. However, it's hardware-dependent, cannot distinguish multiple sources if present, struggles at low SNR, and has a high computation cost. (d) Proposed harmonic-based emanation detection method. A peak detector finds the energy peaks in the spectrum and our computational harmonic detector finds the harmonic patterns. This method addresses all the limitations of the CNN-based method.}
\label{eman_def}
\end{figure}
\subsection{Background}
\begin{figure*}
\centering
\includegraphics[width=7in]{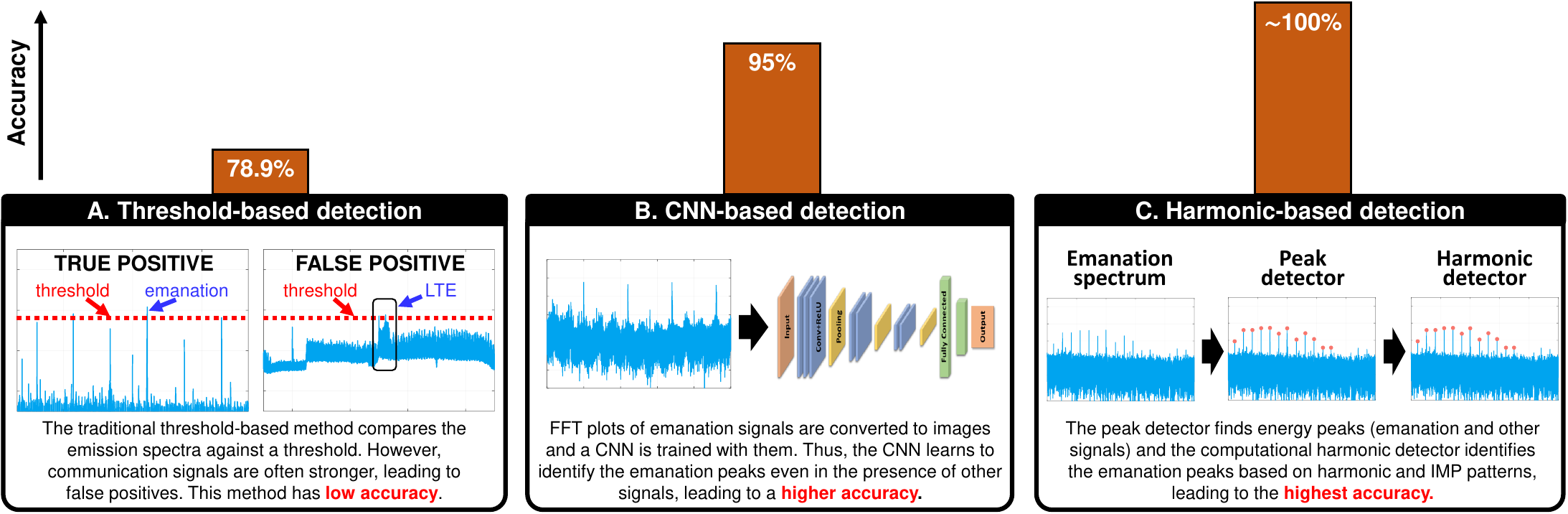}
\caption{Our algorithm development journey from traditional threshold-based method to initial CNN-based method to currently proposed harmonic-based method. (a) Traditionally used threshold-based method compares the emission peaks against a threshold value. However, too high or too low threshold results in false negatives and false positives respectively. In general, this method has the worst performance. (b) In the CNN-based method, FFT plots are treated as images and fed to a CNN which learns to distinguish between the emanation and other signals. This approach renders much higher accuracy, though it has some limitations stemming mostly from the training data. (c) The harmonic-based method exploits emanations' distinct harmonic and IMP patterns compared to other signals. This method performs the best of the three with ${\sim}100\%$ accuracy.}
\label{algos}
\end{figure*}
\IEEEPARstart{D}{igital} signal switching between logic states causes unintended electromagnetic (EM) emissions from electronic devices and connecting wires. This unintentional switching emission is called \textit{`electromagnetic emanation'}. It creates electromagnetic interference (EMI) to the nearby desired signals and may lead to the violation of electromagnetic compatibility (EMC) regulations. In addition to that, the emanation signal contains a significant correlation with the source signal, leading to the recovery of the bit pattern of the source data from it. Essentially emanation provides a \textit{`side-channel for information leakage'} that the attackers can exploit to exfiltrate data, posing a crucial threat to data security. 

Fig.~\ref{eman_def}(a) explains the concept and characteristics of EM emanations. Such EM side-channel leakage has been exploited for cryptographic key recovery \cite{sc0,sc1,sc2}, keystroke inference \cite{kb1,kb2}, monitoring USB device activity \cite{usb_stick}, detecting DNN architecture \cite{dnn_eman}, monitoring smartphone camera activity \cite{smartphone}, cryptographic algorithm and key length detection \cite{crypto_algo}, program activity monitoring \cite{prog_activity}, covert communication \cite{covert1,covert2}, reconstructing screen images \cite{recons_shape, flat_panel, lcd_tv, eaves_mgk}, etc. 
\IEEEpubidadjcol
To secure sensitive facilities, a metal shielding (Faraday Cage) is deployed to contain the emission within it. However, this is not always feasible for government facilities in the wild (e.g., a temporary base in a foreign country). Also, for non-military facilities, shielding is not the best solution because almost all electronic devices leak EM emanation. Shielding the huge number of electronic devices being used every day leads to a massive cost and inconvenience. This is an even more substantial issue for Internet of Things (IoT) devices which are already deployed in large numbers and growing rapidly. Research company IoT Analytics has predicted that there will be nearly 18.8 billion IoT devices by the end of 2024 \cite{iot_count}. Also, even if the shielding is deployed, it may deteriorate over time or be damaged by a malicious adversary, leading to the failure in leakage containment. The best solution is to dynamically monitor the RF spectrum for EM side channel leakage and take steps if significant emanation is detected. However, the current emanation detection method involves RF spectrum sensing using EM probes or SDRs and analyzing them manually. This calls for a research need to develop a smart and automatic emanation detection method which prompted several agencies to fund such research projects (e.g. SCISRS \cite{scisrs} project by IARPA).

Fig.~\ref{eman_def}(b) explains the challenges in detecting EM emanation automatically. Emanations are weak as they are just leakage signals. From the perspective of emanation detection, regular communication signals are much stronger and act as interference. Also, EM emanation is hardware-dependent and its emission frequency can be anywhere in the spectrum. Searching such weak signals within the huge RF spectrum is akin to looking for a needle in a haystack. In our initial observation, it was found that HDMI cables had relatively stronger emanations compared to other off-the-shelf electronic components and peripherals. So, in the preliminary version of this work \cite{date_hdmi}, we collected data from HDMI cables of 3 different shielding types (unshielded, single-shielded, and double-shielded) using Ettus B210 SDR as a receiver (RX) in an office environment. Improving SNR through time and frequency-domain processing and applying transfer learning approach (a pretrained ResNet50 has been retrained using spectrum plots as images), we achieved ${\sim}100\%$ detection accuracy up to \SI{16}{\meter} from the target cable and ${\sim}95\%$ accuracy at \SI{22.5}{\meter}, even in the presence of strong communication signals. This is the highest reported range for HDMI emanation in the literature. Fig.~\ref{eman_def}(c) shows the principle of CNN-based detection. Despite having a high accuracy and detection range, the CNN-based detection approach had some limitations. Firstly, it cannot differentiate between emanations from multiple sources. Additionally, the characteristics of emanations are dependent on the hardware, making it impractical to train a CNN on every possible type of hardware. Consequently, the CNN-based algorithm struggles with new or unfamiliar hardware. Furthermore, the accuracy decreases in low SNR conditions (maximum SNR $<$3 dB) because the fine details often get lost during the conversion from signal to image domain. Lastly, the CNN-based method is computationally intensive due to the complexity of the ResNet50 network, and converting each spectrum to the image domain requires extra computation for the same information content.

In this extended version, we augment our previous study by collecting and characterizing emanation data from a wide range of IoT devices (Arduino, PSoC, ESP32, and Zigbee), other commonly used electronic devices (PC and monitor), and cables (HDMI and USB). Analyzing the data, we find that: (1) the emanation signal consists of energy peaks in the frequency domain with harmonics and intermodulation products or IMP (also forms a harmonic pattern with much smaller frequency separation), (2) the fundamental frequency of the harmonics and frequency separation of intermodulation products (IMP) vary from device to device. While a harmonic detector is an obvious choice to detect the emanations with harmonic patterns, traditional harmonic detectors designed for power systems require apriori knowledge about the fundamental frequency to detect harmonics. Hence, they are infeasible here. Also, frequency separations for intermodulation products are much smaller than fundamental frequency (=harmonic steps) and will require a second harmonic detector. In the field of audio/speech signal processing, there are many works on pitch (=fundamental frequency) estimation, but they do not detect all the other harmonic components. Also, they are not usable for intermodulation products. 

Hence, in this work, we have developed a computational harmonic detection algorithm that doesn't require any apriori knowledge about the fundamental frequency, can detect IMPs simultaneously, and covers a wide range of harmonic patterns found in the experimental data such as regular harmonics, harmonics with one or more missing frequencies, different harmonic groups with 1 common frequency, overlapping IMPs from multiple sources with the same separation, etc. (later, Fig.~\ref{fig_harmonics} shows and section~\ref{sec_var_har_imp} explains these cases in detail). More importantly, the proposed algorithm addresses the limitations of the CNN-based detection method. Fig.~\ref{eman_def}(d) shows the proposed harmonic-based emanation detection method and its advantages. This algorithm is hardware agnostic, can detect multiple emanation sources simultaneously, and can detect low SNR emanations (${\sim}1~dB$). Since it works in the signal domain and there is no signal-to-image conversion and heavy CNN, it is computationally much more efficient. The computational harmonic-based emanation detection method provides ${\sim}100\%$ accuracy up to \SI{22.5}{\meter} for HDMI emanation, compared to ${\sim}95\%$ accuracy achieved in earlier CNN-based methods. Also, it provides ${\sim}100\%$ accuracy for the IoT and electronic devices that we have experimented with. Fig.~\ref{algos} shows our complete journey from the traditional threshold-based method ($78.89\%$) to our initial CNN-based method ($95\%$) and from CNN to our latest harmonic-based detection method (${\sim}100\%$). This manuscript represents $>60\%$ work compared to the conference version \cite{date_hdmi}, especially in the following aspects: (1) extensive emanation data collection from diverse devices, (2) analysis of those data to reveal device-dependent harmonic patterns with intermodulation products, and (3) development of a computational harmonic detection algorithm towards harmonic-based emanation detection method.

\subsection{Our Contribution}
\begin{itemize}
  \item In the preliminary study, using 3 types of HDMI cables (unshielded, single shielded, and double shielded) as target and Ettus B210 SDR as receiver, we have collected HDMI emanation data along with background profiling over 3 days from \SI{0.5}{\meter} to \SI{22.5}{\meter} in an office environment. \textbf{Harnessing the power of DSP techniques to improve the SNR and exploiting the advanced image recognition capability of modern CNN, we have improved the emanation detection range from \SI{4}{\meter} to \SI{22.5}{\meter} for an iso-accuracy of ${\sim}95\%$. Also, ${\sim}100\%$ accuracy is achieved up to a distance of \SI{16}{\meter} from the target.} Comparing the maximum emanation power from HDMI cables with 3 types of shielding, we have evaluated the efficacy of multi-layer shielding for commercially available cables. Also, we have distinguished emanation signals based on their screen content with an accuracy of ${\sim}91.7\%$ at \SI{16}{\meter}.
  \item In this extended version, \textbf{we have collected and characterized emanation data from a wide
  range of IoT devices (Arduino, PSoC, ESP32, and Zigbee), electronic devices (PC and monitor), and cables (HDMI and USB cables).} Our analysis reveals that these emanations have distinct harmonic patterns with intermodulation products.
  \item Based on our study, \textbf{we have developed a computational harmonic detector that is device agnostic, requires no a priori information about the harmonic pattern, can detect low SNR (${\sim}$\SI{1}{\deci\bel}) emanations, covers a wide variety of test cases, and works in the signal domain (unlike CNN-based method, it doesn't require transformation from signal to image).} It provides ${\sim}100\%$ accuracy up to \SI{22.5}{\meter} in the office corridor, compared to ${\sim}95\%$ of the CNN-based method. It also achieves ${\sim}100\%$ detection accuracy for emanations from a wide range of electronic devices and cables.
  \item \textbf{The proposed algorithm has been tested in different environments (anechoic chamber, office building with both line-of-sight and non-line-of-sight cases)} to ensure its efficacy in practical scenarios.
\end{itemize}

\subsection{Organization of the Paper}
The rest of the paper is structured as follows: section~\ref{sec_lit_rev} discusses the relevant works published in the literature. Section~\ref{sec_data_collection} describes our experimental setup and measurement results. This includes both earlier data collection from HDMI for the conference version of this work and new data collection from IoT devices (Arduino, PSoC, Zigbee, and ESP32), other day-to-day electronics (PC, monitor), and cable (USB cable). In Section~\ref{sec_threshold_det}, the traditional threshold-based emanation detection method is explored. Section~\ref{sec_cnn_based_det} describes our initially proposed CNN-based detection method in detail. It also analyzes the effect of multi-layer shielding to suppress emanation. Section~\ref{sec_harmonic_detector} explains the newly proposed harmonic-based emanation detection method. At the heart of this approach, there is a computational harmonic detection algorithm which has been described in detail with pseudocode. Section~\ref{sec_harmonic_emanation} evaluates the performance of the harmonic-based detector, analyzes its device-agnostic detection feature, and compares it with other emanation detection methods proposed in the literature. Also, it is tested in different environments to show its efficacy in various deployment scenarios. Finally, section \ref{conc} concludes the paper with a summary.

\begin{figure*}
\centering
\includegraphics[width=6.6in]{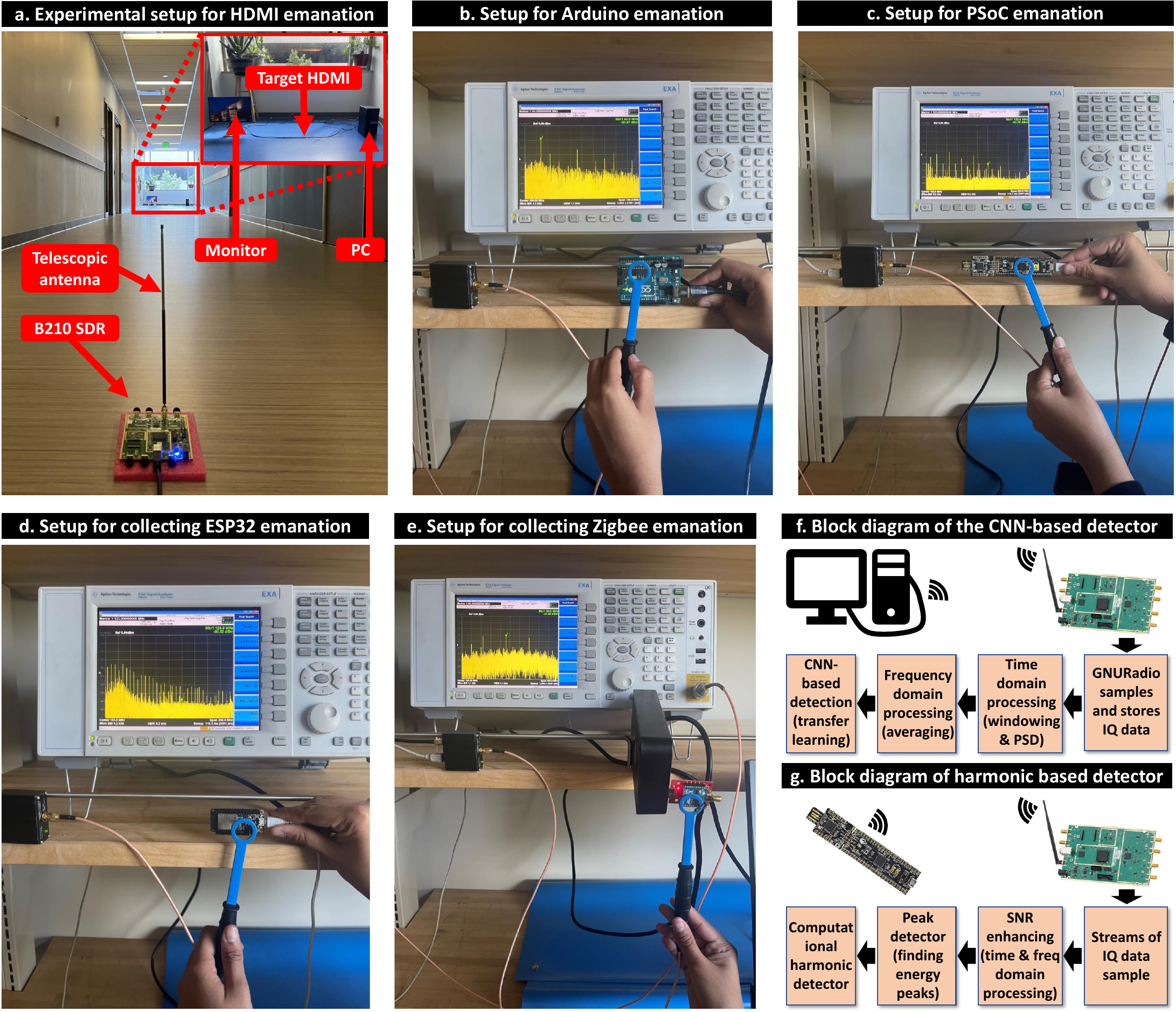}
\caption{(a) HDMI emanation data collection setup in the office corridor using Ettus B210 SDR (connected to a wideband telescopic antenna) as a receiver. The target HDMI is connected between a PC and a monitor (shown in inset). (b) (c) (d) (e) physical setup for emanation data collection from Arduino, PSoC, ESP32, and Zigbee respectively. Emanation is picked up using an H20 magnetic probe, amplified using a 40-dB wideband amplified, and analyzed in a spectrum analyzer. (f) block diagram of the CNN-based emanation detection method that was proposed in the preliminary version of this work \cite{date_hdmi} (g) block diagram of the newly proposed harmonic-based emanation detection method (this work).}
\label{setup}
\end{figure*}

\section{Related Works}
\label{sec_lit_rev}
Unintended electromagnetic leakage or emanation has long been exploited by defense agencies for eavesdropping. During World War II, Bell engineers accidentally noticed such leakage from a 131-B2 mixer that was provided to the Signal Corps by Bell Telephone \cite{tempest0}. With further investigation, the engineers succeeded in recovering 75\% of the plain text of so-called encrypted data, proving the extent of the threat posed by such leakage. After that, emanation was studied by different agencies, leading to many policies and security protocols. The codename \textit{TEMPEST} is used to refer to the classified US government program that studies the `emission security' (EMSEC) issues, possible exploitation, countermeasures, and standardizations (e.g., NATO SDIP-27 Level A, Level B, etc.) \cite{wiki_temp}. Most information regarding TEMPEST is still classified. The first unclassified research work on EM side-channel emission was published by Wim van Eck in 1985 \cite{van_eck}. In a BBC program titled ``Tomorrow's World'', he demonstrated that the screen content can be successfully reconstructed at a long range using very cheap equipment. Electromagnetic emanations are sometimes called `van Eck radiation' after him \cite{wiki_temp}. Repaired cables also show such emanation \cite{bari_broken}. As mentioned in the introduction, electromagnetic emanation is mostly used for various data exfiltration purposes. However, it can also be used for defensive purposes as well. Authors in \cite{eman_finger} have used EM emanation to fingerprint IoT devices. Human-induced EM emanation has been used for access control (touch to unlock) as well \cite{touch_access}. Authors in \cite{ims2022_eman} have utilized such EM leakage to detect the presence of a rogue device in a secure facility. Emanation signal from mobile devices has been proposed to be used for digital forensics \cite{eman_forensic}. Authors in \cite{eman_mb} have used emanation signals to identify motherboard components to find suspicious ICs (probably counterfeited) on board.

The emanation spectrum varies with hardware (even for different programs being executed on the same device). However, a common feature in almost all of them is harmonic patterns and intermodulation products (IMP) forming upper and lower sidebands. So, harmonic detectors can be used to detect their presence. These detectors are widely studied in power systems where they are deployed in the active power filters to detect the harmonics on the power signal created by nonlinear loads and cancel them (by injecting signals with the same magnitude but opposite phase). There are harmonic detectors based on FFT \cite{h_fft1, h_fft2, h_fft3}, wavelets \cite{h_wavelet1, h_wavelet2, h_wavelet3, h_wavelet4}, space vector transformation \cite{h_space}, instantaneous reactive power \cite{h_react1, h_react2}, adaptive variational mode decomposition (AVMD) \cite{h_avmd}, ensemble empirical mode decomposition (EEMD) \cite{h_eemd}, neural networks \cite{h_nn}, etc. A common observation on power system harmonic detectors is that the design and system parameters are selected according to the power grid and fundamental frequency (power supply frequency) which is known beforehand. This approach is not feasible for emanation detection as the fundamental frequency can be anywhere in the RF spectrum.   

There are extensive studies in audio signal processing to estimate the pitch or fundamental frequency of speech or music \cite{pitch1,pitch2,pitch4,pitch5,pitch6,pitch7,pitch8}. However, these algorithms also have some assumptions stemming from speech and music signal properties. Also, they don't detect the full harmonic pattern. Hypothetically, it is possible to estimate the pitch using these audio processing algorithms and then use the power harmonic detectors to find the rest of the harmonics. But none of these are designed for the RF spectrum which is humongous (up to ${\sim}$300 GHz) compared to both audio spectra (usually limited to 20 kHz) and power spectra (50/60 Hz fundamental depending on the country, with harmonics within a few kHz range). A slight error in pitch estimation will lead to missing detection of several harmonics. Let's assume a pitch detector with a 2\% error. So, a 100 MHz fundamental may be estimated as 98 MHz and the harmonic detector will try to find the $10^{th}$ harmonic at 980 MHz, which is 20 MHz off. Considering the bandwidth of many SDRs, this harmonic may not even get detected at all. Finally, these approaches will not work for intermodulation products. To sum it up, a dedicated harmonic detector is required for electromagnetic emanation detection. 

\begin{figure*}
\centering
\includegraphics[width=7in]{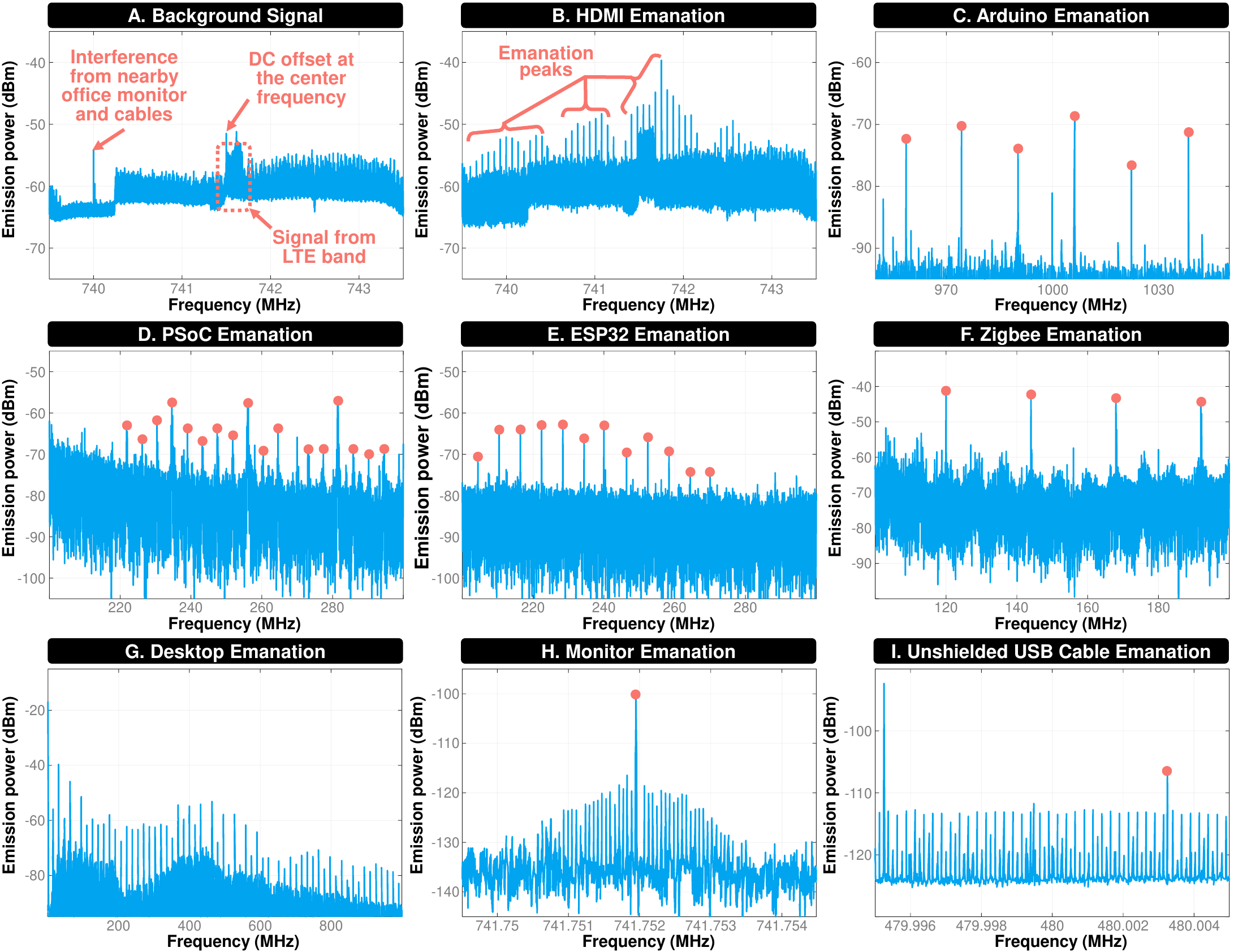}
\caption{(a) RF background signal (HDMI disconnected, PC and monitor off) shows the presence of DC offset, LTE band signal, and other interferer. (b)-(i) Unintended electromagnetic emission spectrum for HDMI, Arduino, PSoC, ESP32, ZigBee, Desktop, Monitor, and Unshielded USB cables respectively.}
\label{eman_spectra}
\end{figure*} 


\begin{table*}[t]
    \centering
    \caption{Description of target equipment for emanation data collection}
    \renewcommand{\arraystretch}{1.4}%
    \begin{tabular}{ | m{8em} | m{8em}| m{42em} | } 
      \hline 
      Equipment type & Target device/cable & Description \\
      \hline \hline
      \multirow{4}{8em}{IoT Devices} & Arduino & It's a microcontroller board used in many academic projects, robotics, home automation, low-cost scientific instruments, etc. Our specific board is Arduino Uno R3 with ATMega328P MCU \cite{arduino}. \\\cline{2-3}
       & PSoC & It's a family of programmable embedded systems based on the ARM Cortex-M processor. PSoC is used in industrial automation, household appliances, medical devices, security systems, etc. Our specific model is 32-bit PSoC 5LP Arm Cortex-M3 \cite{psoc} with CY8C58LP family SoC. \\\cline{2-3} 
       & Zigbee & These devices are used to create wireless mesh networks for building automation, lighting, smart city, medical, and asset tracking \cite{zigbee}. It is based on IEEE 802.15.4 specifications. Our specific model is Xbee S2C. \\\cline{2-3}
      & ESP32 & It is a series of low-cost, low-power microcontroller modules with Wi-Fi and Bluetooth connectivity. They are used for a wide range of applications including smart industrial devices, PLCs, smart medical devices, smart energy devices (HVAC, thermostats, etc.), wearable health monitors, etc. We use ESP32-DevKitC with ESP32-WROOM-32E module on board \cite{esp32}. \\ 
      \hline
      \multirow{2}{8em}{Other Electronic Devices} & Desktop Computer & We have used a Dell OptiPlex PC with an Intel core\texttrademark~i7-6700 processor and 8 GB of RAM (2400 MHz) \\\cline{2-3}
       & Monitor & We have used a Dell P2319H monitor which is a 1080p 60Hz LED monitor. \\
      \hline
       \multirow{2}{8em}{Cables} & USB Cable & USB cables are most commonly used to connect USB peripherals. We have used an unshielded USB cable. \\\cline{2-3}
       & HDMI Cable & HDMI is the most common display cable nowadays \cite{why_hdmi1}. In our preliminary version of this work \cite{date_hdmi}, we used 3 types of HDMI 2.0 cables (shielded, single-shielded, and double-shielded). \\
      \hline
    \end{tabular}
    \label{tab1}
\end{table*}

\section{Data Collection}
\label{sec_data_collection}
\subsection{Target Equipment}
For emanation data collection, the target devices are chosen in such a way that a diverse set of electronic devices with various microcontrollers and system-on-chips (SoC) are included. It is impossible to cover all types of leaking electronics, but with careful selection, the collected data can be representative of the majority of the electronic devices being used every day without the loss of generality. With that in mind, data were collected from 3 groups of devices (IoT, everyday electronics, and commonly used cables) which are described in Table~\ref{tab1}.

\subsection{Experimental Setup}
Fig.~\ref{setup}(a) shows our experimental setup for emanation data collection from an HDMI cable in an office corridor (the inset shows a zoomed-in view of the target HDMI). An Ettus B210 SDR connected with a wideband telescopic antenna is used as a receiver. A GNURadio interface collects the received data, samples at \SI{4}{\mega S \per\sec}, and stores them. Fig.~\ref{setup}(b)-(e) shows our experimental setup for Arduino, PSoC, ESP32, and Zigbee respectively. Emanation signals are picked up by an EM probe (H20), amplified by a 40-dB wideband LNA, and fed to a spectrum analyzer. 

Fig.~\ref{setup}(f) shows the block diagram of the CNN-based emanation detection method. Captured samples are processed in the time and frequency domain to improve SNR and converted to images (spectrum plots). These plots are then fed to a CNN to detect the emanation signal. Fig.~\ref{setup}(g) shows the block diagram of the newly proposed harmonic-based emanation detection method. Captured IQ data are processed to improve SNR. But then instead of converting to spectrum images, energy peaks are detected using a peak detector. The peaks are fed to our computational harmonic detector to find the harmonic group(s) and corresponding emanation signal(s), if any.

\subsection{Measurement Results and Analysis}
\begin{figure*}
\centering
\includegraphics[width=6.8in]{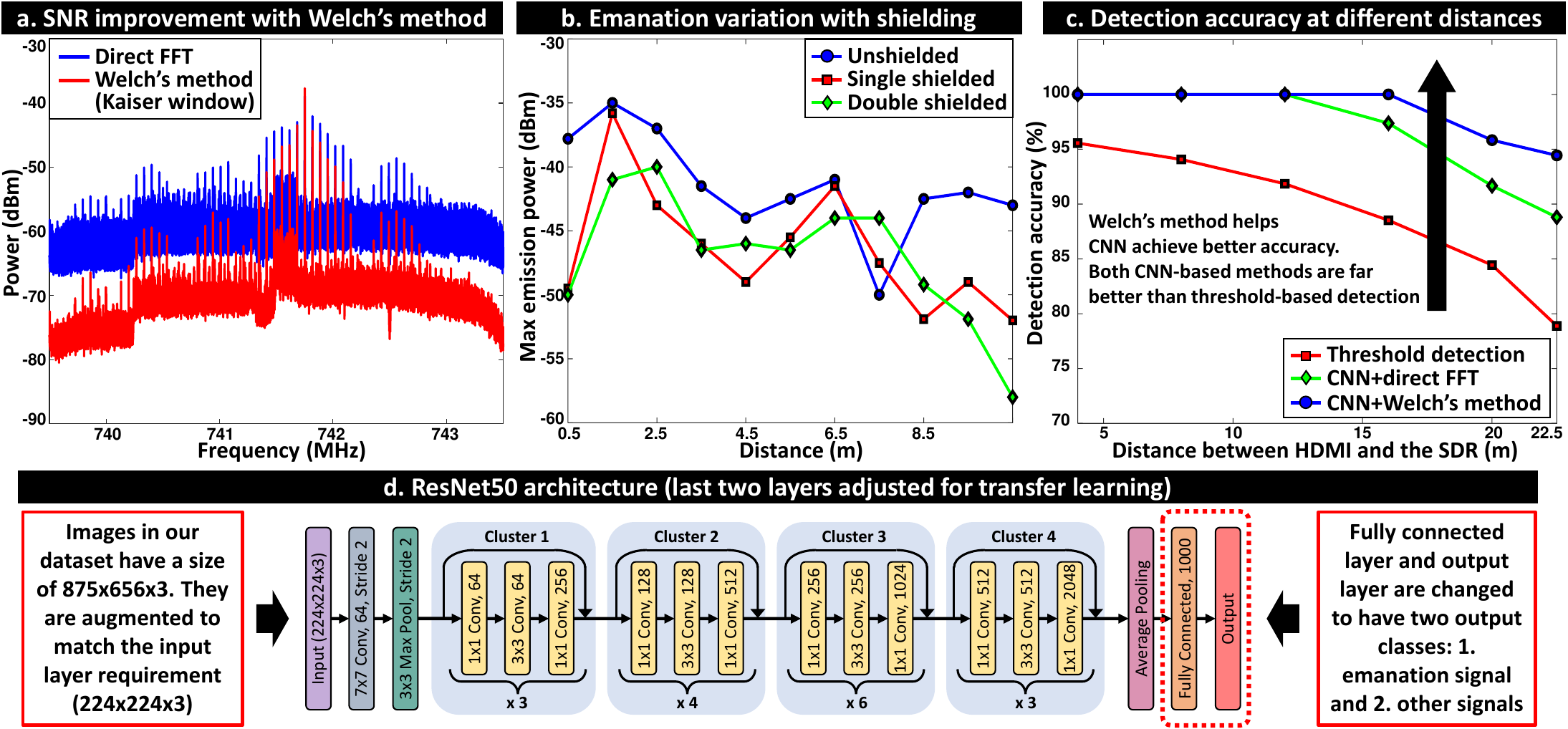}
\caption{(a) Using Welch's method of power spectrum estimation (along with Kaiser window), we get ${\sim}\SI{15}{\deci\bel}$ SNR improvement. (b) Comparison of 3 types of HDMI (unshielded, single-shielded, and double-shielded) in terms of maximum emanation power. The unshielded cable is much louder in most cases while the difference between the other two is trivial. (c) Performance comparison between threshold-based detection (in the best case at -55 dBm threshold) and CNN-based detection (both for direct FFT and Welch's method). SNR improvement provides a better detection range for iso-accuracy and a better accuracy for iso-distance than FFT from raw data. In both cases, the CNN-based method outperforms the threshold-based method by a high margin. (d) ResNet50 network which is used in the CNN-based method. Images in our dataset are augmented to match the input size (224x224x3). The fully connected layer and the output layer are modified to classify two groups: emanation vs every other signal.}
\label{snr_imp}
\end{figure*}
\subsubsection{Background Measurement}
For the CNN-based method (preliminary version of this work), HDMI emanation was considered as positive class, while RF background data (HDMI disconnected and there is no HDMI emanation) was labeled as negative class. To make the background profile robust, we collected background data on 3 separate days at 3 separate times: in the morning, at noon, and at night. Fig.~\ref{eman_spectra}(a) shows a sample background spectrum. We chose \SI{742.5}{\mega\hertz} or the $5^{th}$ harmonic as the target frequency where HDMI emanation was observed to be the strongest. Coincidentally, it overlaps with the LTE band (lower SMH block, from \SI{729}{\mega\hertz} to \SI{746}{\mega\hertz}). Fig.~\ref{eman_spectra}(a) clearly shows the LTE signal energy. Also, there are multiple experimental labs and office rooms on both sides of the corridor, contributing to some additional interference. Furthermore, the DC offset can be seen at the center. For the positive class, HDMI emanation data were collected from \SI{0.5}{\meter} to \SI{22.5}{\meter}, at \SI{0.5}{\meter} intervals for 3 types of cables: unshielded, single-shielded, and double-shielded.  

\subsubsection{Spectral Analysis}
Fig.~\ref{eman_spectra}(B)\textendash(I) shows emanation spectra from HDMI, Arduino, PSoC, ESP32, Zigbee, PC, monitor, and USB cables respectively. The key observation is that each device has its distinct electromagnetic leakage with unique harmonic patterns and sidebands (consisting of intermodulation products or IMP). These patterns can be utilized to identify them uniquely. Table~\ref{table_spectrum} lists the fundamental frequency and intermodulation product values for each device.

\begin{table}[h!]
\centering
\caption{Harmonic and IMP patterns of test devices}
\begin{tabular}{||c | c | c ||} 
 \hline
 Emanation source & Fundamental frequency (MHz) & IMP step (MHz) \\
 \hline
 HDMI & 148.5 & 0.07 \\ 
 \hline
 Arduino & 16 & many IMPs \\
 \hline
 PSoC & 64 & 1.6 \\
 \hline
 ESP32 & 6 & 0.525 \\ 
 \hline
 ZigBee & 24 & 3 \\
 \hline
 Desktop & 3.3 & - \\
 \hline
 Monitor & 148.5 & 0.07 \\ 
 \hline
 USB & 480 & 0.00025 \\
 \hline
\end{tabular}
\label{table_spectrum}
\end{table}

\section{Traditional Threshold-Based Detection}
\label{sec_threshold_det}
\subsection{Detection Method}
Threshold-based peak detection is widely used in literature, including some recent ones \cite{ims2022_eman}. The collected I-Q data are transformed into the frequency domain via FFT and compared against a threshold level. If there is a power peak above the threshold, we detect the presence of an emanation signal.

\begin{table}[h!]
\centering
\caption{Performance analysis of threshold-based detection}
\begin{tabular}{||c | c | c | c||} 
 \hline
 Threshold (dBm) & FP (\%) & FN (\%) & Accuracy (\%) \\
 \hline
 -45 & 0 & 74.07 & 62.96 \\ 
 \hline
 -55 & 26.67 & 15.56 & 78.89 \\
 \hline
 -65 & 100 & 0 & 50 \\
 \hline
\end{tabular}
\label{table_1}
\end{table}

However, the accuracy is dependent on the threshold. If the threshold is too high, there will be a lot of false negative (FN) values, whereas too-low thresholds result in a lot of false positives (FP). We have tested the detection performance of our dataset for 3 threshold levels. Table~\ref{table_1} shows that the \SI{-55}{\deci\bel\meter} threshold provides the best accuracy, which is still pretty low (78.89\%).

\subsection{Challenges}
Table~\ref{table_1} shows that threshold-based detection does not perform well (best-case accuracy $<80\%$). There are several reasons for that. The key issue is the strong interference from other signals (LTE signals, emanations from other cables, etc.). These peaks in the background are falsely detected as emanations for low thresholds, leading to high FP values. On the other hand, raising the threshold does not help much as emanation signals are weak and cannot cross the high threshold bar. This leads to high FN values. Also, the background noise level keeps changing based on the environment. This poses a challenge for adaptive threshold selection. Another issue is the peak at the center frequency due to DC offset. To address these issues, we resort to CNN-based detection.

\section{CNN-Based Detection}
\label{sec_cnn_based_det}
\subsection{SNR Improvement using DSP Techniques}
\label{subsec_snr}
Before training a CNN, we want to improve the perceived SNR of the collected signal. To that end, we apply some known techniques in the DSP domain.
\subsubsection{Windowing in Time Domain}
Our data is finite in the time domain, which is equivalent to applying a rectangular window to an infinite time sequence of data. However, the rectangular window has a significant spectral leakage from the main lobe to the side lobe \cite{dsp_book}. There are better windows (Hanning, Hamming, Kaiser, etc.) with lower spectral leakage. The downside is the larger main lobe width. However, the main lobe width is inversely proportional to the data length and we have significantly long data to overcome this limitation. The spectral leakage for the Kaiser window ($\beta = 5.66$) is only $0.01\%$ compared to $9.28\%$ for the rectangular window. Also, relative side-lobe attenuation reduces from \SI{-13.3}{\deci\bel} to \SI{-41.4}{\deci\bel}. We have windowed our data with the designed Kaiser window ($\beta = 5.66$).

\subsubsection{Power Spectrum Estimation using Welch's Method}
We have used Welch's method of power spectrum estimation (modified periodogram with averaging) as a better spectrum estimate. A sequence of $4\times10^5$ samples (\SI{0.1}{\second} data) is taken and divided into $8$ segments with $50\%$ overlap. A modified periodogram (FFT of autocorrelation, instead of direct FFT) is applied to each segment and the output is averaged. Fig.~\ref{snr_imp}(a) compares the spectrum of I-Q data corresponding to emanations from unshielded HDMI cable at \SI{1}{\meter} distance, found using direct FFT and Welch's method. It is shown that the maximum peak is similar for both, but the noise level is reduced. We gain $\sim \SI{15}{\deci\bel}$ SNR improvement.

\subsubsection{Averaging in Frequency Domain}
Frequency domain averaging reduces noise power, keeping the signal peaks almost intact. We apply Welch's method for $9$ consecutive sequences with $50\%$ overlap in the time domain. The spectra of the sequences are averaged to reduce the noise power.

\subsection{Effect of Multi-Layer Shielding on Emanation}
After improving the SNR, we want to check the efficacy of multi-layer shielding before moving on to training CNN. Fig.~\ref{snr_imp}(b) shows the emanation power for $3$ types of HDMI cables from \SI{0.5}{\meter} to \SI{10.5}{\meter} at \SI{1}{\meter} intervals. Except for a few outliers, the maximum emanation power for the unshielded cable (blue line) is significantly higher than the other two, which is expected. However, the difference between the single and double-shielded cables is trivial. 
\begin{figure*}
\centering
\includegraphics[width=6.6in]{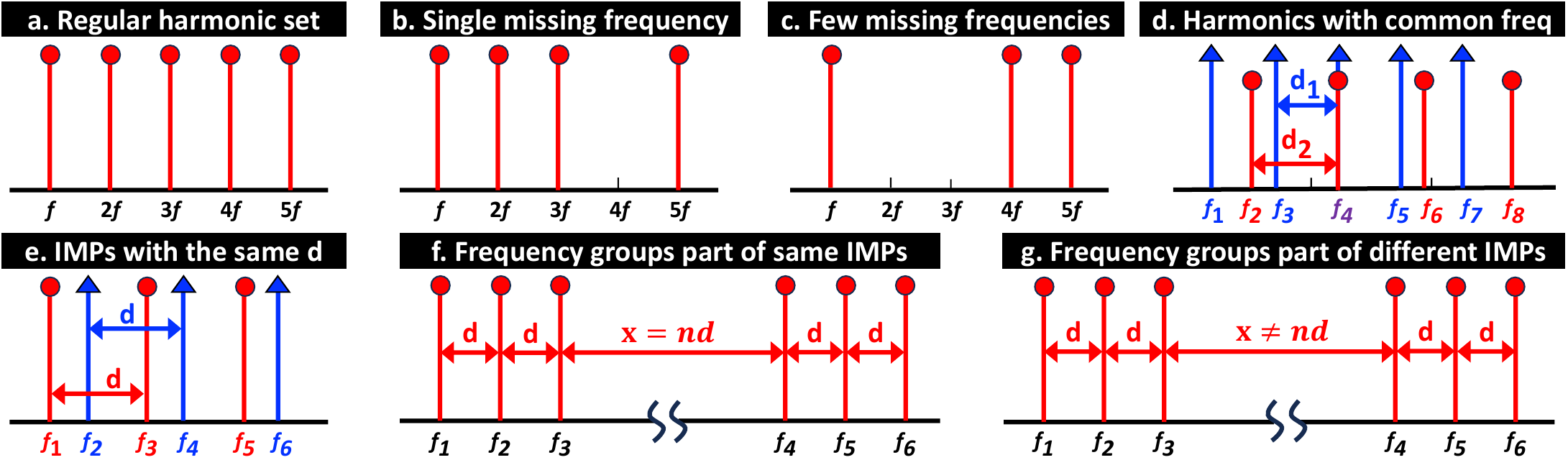}
\caption{(a) Regular harmonic frequencies with fundamental frequency $f$. (b) Missing single harmonic (here, $4^{th}$ harmonic is missing). (c) Missing multiple harmonic frequencies (here, $2^{nd}$ and $3^{rd}$ harmonics are missing). (d) Two harmonic groups have a common frequency ($f_4$) in them. (e),(f),(g) are more specific to IMPs only. (e) Two overlapping IMP groups with same frequency separation $d$. (f) Two separate IMP groups with same frequency separation $d$, but far away from each other (group separation $x=nd$, where n is large. Essentially, they are part of same IMP group. (g) Similar to the previous case, two separate IMP groups with same frequency separation $d$, but far away from each other. However, group separation $x \neq nd$, where n is large. Hence, they are two distinct IMP groups.}
\label{fig_harmonics}
\end{figure*}
\subsection{Transfer Learning using ResNet50}
Our final step is to evaluate the performance of the CNN-based detection method. For that, we have used transfer learning approach where a widely used, pretrained CNN (e.g. VGG16 \cite{vgg16}, AlexNet \cite{alex}, GoogLeNet \cite{google}, ResNet50 \cite{res50}, etc.) is retrained \cite{xfr, xfr2} with a new dataset. These networks are carefully designed and reviewed by experts in the field and are known to classify images in standard datasets (e.g. ImageNet) with high accuracy. Our initial testing shows that ResNet50 works best for our case.

Residual Neural Network or ResNet revolutionized the use of the ultra-deep network by using `skip connection' to address the issue of `vanishing gradient' and `degradation problem'. It performed much better compared to VGG or GoogLeNet on the `ImageNet dataset' \cite{resnet_intuition}. We will exploit the enhanced image classification capability of ResNet50 for our dataset. For training purposes, we have used the frequency domain plots as images. The rationale behind using the plots instead of the 1D sequence is to exploit the advanced image classification capability of CNN (ResNet50). Our images are augmented to match the input size. The fully connected layer and the output layer are adjusted for binary classification (emanation vs others). The train, test, and validation data ratio was 70:20:10. We have used an initial learning rate = 0.001, mini-batch size = 8, and max epoch = 30. Data are shuffled at each epoch.

\subsection{Performance Evaluation}
Fig.~\ref{snr_imp}(c) shows the performance of ResNet50 via transfer learning for a distance of \SI{4}{\meter} to \SI{22.5}{\meter}. For data with better SNR (thanks to Welch's method), we get $\sim100\%$ accuracy up to \SI{16}{\meter} which gradually reduces to $\sim95\%$ at \SI{22.5}{\meter}. This figure also compares the performance benefit with improved SNR. Compared to direct FFT, we get a longer distance for iso-accuracy (e.g. for $100\%$ accuracy, we get \SI{16}{\meter} compared to \SI{12}{\meter}) and higher accuracy for the same distance (e.g. at \SI{22.5}{\meter}, we get $\sim95\%$ accuracy compared to $\sim88.9\%$). In both cases, the CNN-based method outperforms the threshold-based method by a high margin.


\subsection{Monitor Content Type Detection - Still vs Video}
In the literature review section, we have discussed that some works have reconstructed screen images with plain text and some geometric shapes. Some recent works have tried to reconstruct video signals. However, these works assume that the monitor is running solely a video or just an image. An automated method is necessary to detect the content type of the monitor (still image vs video) and switch to the corresponding detection algorithm. In this subsection, we try to fill that void. 

Our data collection method is the same as before, except that a video was playing on the monitor. The emanation signal captures the changes due to the video playback. We have used the same data processing steps before training ResNet50 with the data to check whether it can distinguish between the two. Our evaluation shows that we get $91.7\%$ accuracy at a distance of \SI{16}{\meter} which gradually drops at a further distance.

\section{Harmonic-based Emanation Detection}
\label{sec_harmonic_detector}
\subsection{Harmonic and Intermodulation Product (IMP) Types}
\label{sec_var_har_imp}
\begin{figure*}
\centering
\includegraphics[width=6.8in]{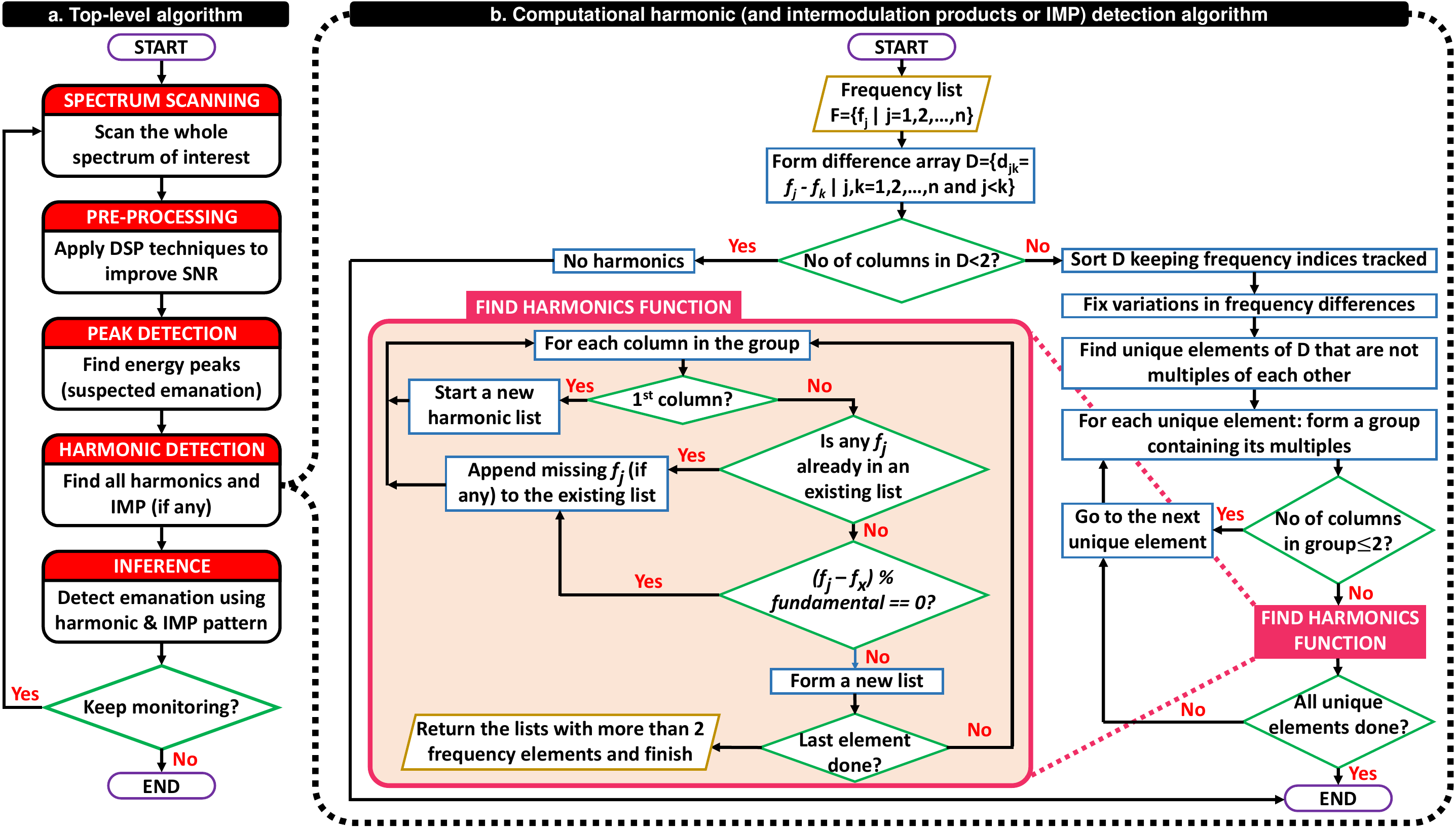}
\caption{(a) Flow diagram of the harmonic-based emanation detection algorithm. At the $4^{th}$ step, it requires harmonics and intermodulation products detection, where our computational harmonic detector comes into play. (b) Logical flow diagram of the computational harmonic detector.}
\label{fig_flowchart}
\end{figure*}
Fig.~\ref{fig_harmonics} shows various harmonic and IMP patterns. Fig.~\ref{fig_harmonics}(a) shows a group of regular harmonic frequencies where the fundamental frequency is $f$. However, in practical scenarios, often 1 or 2 harmonics are missing. Fig.~\ref{fig_harmonics}(b) shows $4^{th}$ harmonic missing and Fig.~\ref{fig_harmonics}(c) shows both $2^{nd}$ and $3^{rd}$ harmonics are missing. A little complicated scenario is shown in Fig.~\ref{fig_harmonics}(d) where we have two overlapping harmonic groups marked by blue (consisting of frequencies $f_1$, $f_3$, $f_4$, $f_5$, and $f_7$) and red color (consisting of frequencies $f_2$, $f_4$, $f_6$, and $f_8$). Both have a frequency $f_4$ common in them. Fig.~\ref{fig_harmonics}(e), (f), and (g) show several cases that are mainly found for IMPs. Fig.~\ref{fig_harmonics}(e) shows two different IMP groups having the same frequency difference $d$. Fig.~\ref{fig_harmonics}(f) shows two IMP groups with the same frequency separation $d$, but far away from each other. The separation between the groups is $x=nd$, where n is a large number. Hence, they are essentially the same IMP group. A contrasting scenario to this is shown in Fig.~\ref{fig_harmonics}(g). Here, we also have two IMP groups with separation $x\neq nd$. Hence, they are distinct IMPs from different sources. Our algorithm is developed to cover all these cases.

\subsection{Top Level Algorithm}
Fig.~\ref{fig_flowchart}(a) shows the top-level abstract of the harmonic-based emanation detection algorithm. First, the spectrum of interest is scanned. If the spectrum is enormous, it is divided into smaller chunks of uniform frequency span to ensure good enough frequency resolution to distinguish intermodulation products (IMP). Next, the scanned data are processed using DSP techniques described in subsection~\ref{subsec_snr} to improve SNR. Then, the spectrum data are fed into a peak detector that detects all the energy peaks in the frequency domain. These are suspected emanations. In the next step, the list of peak frequencies is fed into our computational harmonic detector, which finds all the harmonics and intermodulation products. The `HARMONIC DETECTION' step from Fig.~\ref{fig_flowchart}(a) has been expanded into Fig.~\ref{fig_flowchart}(b) to show the flow diagram of the underlying computational harmonic detector algorithm. This algorithm is explained later in subsection~\ref{subsec_CHD}. Finally, by analyzing the detected harmonic and IMP pattern, it is inferred whether any unintended emission is present. 

\subsection{Peak Detection}
Emanations appear as spurious peaks in the frequency domain. A simple peak detector can detect those peaks and send that frequency list to the computational harmonic detector. The wavelet transform-based peak detector \cite{cwt_peak_detector} is widely used and available as a built-in function in the SciPy Python module \cite{scipy_cwt}. This built-in function is used in our code.

\subsection{Helper Functions Development}
To keep the main algorithm organized and modular, two helper functions are developed. The first helper function is a custom quicksort function that keeps track of the original indices. Let's assume we have a $3 \times n$ array whose first row contains pixel values while the second and third rows contain its coordinates or $(x,y)$ indices. We want to sort the pixel values while keeping their original indices with them. To achieve this, a modified version of the well-known quicksort algorithm is implemented where the comparison is made based on $1^{st}$ row values, but during binary grouping, the whole column is appended. Algorithm~\ref{alg:alg1} shows this custom quicksort method.

In the experimental data, harmonic values are often not exact multiples of each other. Rather, the harmonics have slightly different frequencies than expected. To address this, a second helper function is defined which takes a tolerance limit and finds the group of values that differ within that tolerance range. Then it replaces the whole group with the average of it. Algorithm~\ref{alg:alg2} shows the pseudocode of this function.

\begin{algorithm}[!htb]
\caption{Customized Quicksort Algorithm}
\label{alg:alg1}
\begin{algorithmic}[1]
\Function{{custom\_quicksort}}{$X$}
\State $r \gets \text{no of rows in X}$
\State $c \gets \text{no of columns in X}$
\If{$c<2$}
\State return X
\EndIf
\State $X1, X2 \gets [~]$
\For{$n = 0, \dots, c-2$}
\If{$X[0][i] \leq X[0][c-1]$}
\State $append(X1, X[:,i])$
\Else
\State $append(X2, X[:,i])$
\EndIf
\EndFor
\State return $append(CUSTOM\_QUICKSORT(X1),\\X[:,c-1],CUSTOM\_QUICKSORT(X2))$
\EndFunction
\end{algorithmic}
\end{algorithm}

\begin{algorithm}[!htb]
\caption{Function to Fix Differences in Harmonic Steps}
\label{alg:alg2}
\begin{algorithmic}[1]
\Function{fix\_freq\_var}{$D, D_{index}, \epsilon_{freq}$}
\State $i \gets 0$
\While{i $<D_{index}-1$}
\State $j \gets i$
\While{$|D[0][i] - D[0][j+1]|\leq \epsilon_{freq}\times D[0][j+1]$}
\State $j \gets j+1$
\If{j==$D_{index}-1$}
break
\EndIf
\EndWhile
\If{j$>$i}
\State $D[0][i:j+1] \gets average(D[0][i:j+1])$
\EndIf
\State $i \gets j+1$
\EndWhile
\State return \textit{D}
\EndFunction
\end{algorithmic}
\end{algorithm}

\subsection{Computational Harmonic Detector}
\label{subsec_CHD}
Fig.~\ref{fig_flowchart}(b) shows the logical flow diagram of the algorithm and Algorithm~\ref{alg:alg3} shows its pseudocode. It works in the following steps:

\textbf{Step1:} A list of frequencies $F=\{f_j~|~j=1,2,...,n\}$ is taken as the input. From this frequency list, differences between every possible pair of frequencies are calculated ($d_{jk}=f_j - f_k~|~j,k=1,2,...,n$) to form a difference matrix $D$. However, such a matrix will be symmetric, with the principal diagonal being all `0'. Hence, for computational efficiency, $d_{jk}$ is calculated only for $j<k$. The values are stored as a $3\times n$ array with differential values ($d_{jk}$) in the first row and the frequency indices ($j,k$) in the second and third rows. Now, if there is only 1 column in the difference array, then there are no harmonics and the process is terminated. Otherwise, the algorithm moves on to the next step.
    
\textbf{Step2:} In this step, the helper functions that were developed before are used. First, the difference array is sorted using our custom quicksort function, keeping frequency indices tracked. Next, for the harmonic detection case (not IMP), the slight variation in difference values is fixed using our helper function. 

\textbf{Step3:} Certain $d_{jk}$ values are unique, while other values are multiples of these. Next, these unique difference values are found. For each unique value, its integer products are taken with it to form a product group. From each of these groups, one or more harmonic sets are found using the FIND\_HARMONICS function. But before calling that function, certain groups are ignored.

Assume $3$ frequencies ($f_1$, $f_2$, and $f_3$) with $2$ difference values: $d_{12}=d$ and $d_{23}=Nd$, where $N \in Z$. However, there must be another differential value $d_{13}=(N+1)d$ in the group. In other words, there cannot be only $2$ values in the product group in practical scenarios. Groups with $\leq2$ elements are created numerically, while they cannot exist experimentally. Hence, such groups are discarded. Also, specifically for IMPs, there might be some groups with $>2$ elements while all of them are the same. Let's assume ($N+1, N \in Z$) IMP frequencies with $N$ number of difference values, all being $d$. It's apparent that such $N$ element group can not exist in reality as there must be at least another element with a differential value $Nd$. Hence, such groups are discarded.

\textbf{Step4:} For each group formed in the previous step, FIND\_HARMONICS function is called. Its pseudocode is given in Algorithm~\ref{alg:alg4}. For the first column of the group, a new harmonic list is formed with the $2$ frequency indices of that column. For all other columns, if any (or both) of the frequency indices are already in an existing list, it's simply part of that harmonic or IMP group. If only one frequency index matches, the other frequency index is appended to that list. But what happens if none of the indices are part of any existing harmonic lists? That can never happen for harmonics because $2$ harmonic groups with the same difference (=fundamental frequency) are basically the same harmonic group. 

    
However, this condition may arise for IMPs with $2$ possible scenarios as was described in Fig.~\ref{fig_harmonics}(f) and Fig.~\ref{fig_harmonics}(g). Firstly, if the difference between any frequency in one of the existing lists and the test frequency indices is a multiple of the fundamental step of that list, the $2$ frequency indices of the test column are part of it (similar to Fig.~\ref{fig_harmonics}(f)). So the test column's indices are appended to that list. However, if the difference is not a multiple of the fundamental step, the test column is part of a new IMP group as was shown in Fig.~\ref{fig_harmonics}(g). So, a new harmonic list is formed with those two frequency indices. The loop keeps running till the last column of the group. In the end, it discards any harmonic list with only $2$ frequency elements and returns the other harmonic lists. In the main harmonic detector function, harmonic lists returned by the FIND\_HARMONICS function (for each unique element group) are appended together to form a final harmonic or IMP list.

\begin{algorithm*}[!htb]
\caption{Harmonic and Intermodulation Products Detection Algorithm}
\label{alg:alg3}
\begin{algorithmic}[1]
\LComment{Initializing variables and arrays}
\State $Freq\_list \gets \text{Frequencies where peaks are found}$
\State $D_{min} \gets \text{Minimum difference}$
\State $D_{max} \gets \text{Bandwidth/2}$
\State $\epsilon_{freq} \gets \text{Tolerance for frequency variation}$
\Statex
\Function{Detector}{$Freq\_list, D_{min}, D_{max}, \epsilon_{freq}$}
\State $harmonic\_freqz \gets \{~\}$ \Comment{Empty Python Dictionary}
\State $harmonic\_steps \gets [~]$
\State $unique\_steps \gets [~]$
\State $D \gets [~]$
\State $n\_harmonic\_group, D_{index} \gets 0$
\LComment{Difference matrix D formation in a $(3\times n)$ array}
\For{$i \textbf{ in } range(len(Freq\_list))$}
\For{$j \textbf{ in } range(len(Freq\_list))$}
\State $difference=|Freq\_list(j) - Freq\_list(i)|$
\If{i$\geq$j \textbf{or} $difference<D_{min}$ \textbf{or} $difference>D_{max}$}
\State continue
\Else
\State append(\textbf{D}, [difference, i, j])
\State $D_{index} \mathrel{{+}{=}} 1$
\EndIf
\EndFor
\EndFor
\LComment{If D has 1 column, return empty array and dictionary as there are no harmonics}
\If{$D_{index} < 2$}
\State \textbf{return} \textit{harmonic\_steps, harmonic\_freqz}
\EndIf
\LComment{Sorting difference matrix columns based on difference values (first row) from small to large}
\State $D \gets CUSTOM\_QUICKSORT(D)$
\Comment{CUSTOM\_QUICKSORT() function is defined in algorithm~\ref{alg:alg2}}
\If{${\sim}flag\_subband$}
\State $D \gets FIX\_FREQ\_VAR(D, D_{index}, \epsilon_{freq})$
\Comment{FIX\_FREQ\_VAR() function is defined in algorithm~\ref{alg:alg3}}
\EndIf
\Statex
\LComment{Find harmonics or intermodulation products}
\For{$i \textbf{ in } range(len(D[0]))$}
\If{i == 0}
\State $append(unique\_steps, D[0,0])$
\State $group \gets D[:,(D[0]~\%~D[0][0])==0]$
\ElsIf{$(D[0,i]~\%~unique\_steps)==0$}
\State $append(unique\_steps, D[0,i])$
\State $group \gets D[:,(D[0]~\%~D[0][i])==0]$
\Else
\State continue
\EndIf
\If{$len(group)\leq2$}
\State continue
\ElsIf{$flag\_subband$ \textbf{and} $len(set(group))<2$}
\State continue
\Else
\LComment{FIND\_HARMONICS() function is defined in algorithm~\ref{alg:alg4}}
\State $temp\_harmonic \gets
FIND\_HARMONICS(group, Freq\_list)$
\State $n\_temp\_harmonic \gets len(temp\_harmonic)$
\If{$n\_temp\_harmonic==0$}
continue
\Else
\State append$(harmonic\_steps, group[0][0]*n\_temp\_harmonic)$
\For{$(j = n\_harmonic\_group, \dots, n\_harmonic\_group + n\_temp\_harmonic)$}
\State harmonic\_freqz[j]=sort(temp\_harmonic[j-n\_harmonic\_group])
\EndFor
\State $n\_harmonic\_group \mathrel{{+}{=}} n\_temp\_harmonic$
\EndIf
\EndIf
\EndFor
\State \textbf{return} $harmonic\_steps, harmonic\_freqz$
\EndFunction
\end{algorithmic}
\end{algorithm*}

\begin{algorithm*}[!htb]
\caption{FIND\_HARMONICS Function to Find IMPs of Same Differential Values and Harmonics}
\label{alg:alg4}
\begin{algorithmic}[1]
\Function{find\_harmonics}{$group, Freq\_list$}
\State $harmonic\_subfreqz \gets \{~\}$
\State $n\_harmonic \gets 0$
\State $FLAG=False$
\For{$i \textbf{ in } range(len(group[0]))$}
\If{$i==0$}
\State $harmonic\_subfreqz[n\_harmonic]=[group[1][0],group[2][0]]$
\State $n\_harmonic \mathrel{{+}{=}} 1$
\Else
\For{$j \textbf{ in } range(n\_harmonic)$}
\State $flag\_match\_first \gets (harmonic\_subfreqz[j]==group[1][i])$
\State $flag\_match\_second \gets (harmonic\_subfreqz[j]==group[2][i])$
\If{$flag\_match\_first==True$ \textbf{and} $flag\_match\_second==True$}
\State $FLAG=True$
\State $break$
\ElsIf{$flag\_match\_first==True$ \textbf{and} $flag\_match\_second==False$}
\State $FLAG=True$
\State $append(harmonic\_subfreqz[j], group[2][i])$
\State $break$
\ElsIf{$flag\_match\_first==False$ \textbf{and} $flag\_match\_second==True$}
\State $FLAG=True$
\State $append(harmonic\_subfreqz[j], group[1][i])$
\State $break$
\Else
\State $separation \gets Freq\_list[harmonic\_subfreqz[j][-1]] - Freq\_list[group[1][i]]$
\If{$(separation~\%~group[0][0])==0$}
\State $FLAG=True$
\State $append(harmonic\_subfreqz[j], [group[1][i],group[2][i]])$
\State $break$
\Else
\State $continue$
\EndIf
\EndIf
\EndFor
\If{$FLAG==True$}
\State $FLAG=False$
\Else
\State $harmonic\_subfreqz[n\_harmonic]=[group[1][i],group[2][i]]$
\State $n\_harmonic \mathrel{{+}{=}} 1$
\EndIf
\EndIf
\EndFor
\State $Final\_Dict \gets \{~\}$
\State $j \gets 0$
\For{$i \textbf{ in } range(n\_harmonic)$}
\If{$len(harmonic\_subfreqz[i])>2$}
\State $Final\_Dict[j] = harmonic\_subfreqz[i]$
\State $j \mathrel{{+}{=}} 1$
\Else
\State $continue$
\EndIf
\EndFor
\State return $Final\_Dict$
\EndFunction
\end{algorithmic}
\end{algorithm*}


\section{Results and Discussion}
\label{sec_harmonic_emanation}
\subsection{Performance Evaluation of Harmonic-based Detector}
Applying our harmonic-based emanation detection method, we achieve ${\sim}100\%$ accuracy for all the test devices. However, the distance up to which this accuracy is achieved varies from device to device. Hence, accuracy alone doesn't paint the full picture. Both the accuracy and the distance at which that accuracy is achieved are required for the complete performance evaluation. Table~\ref{table_max_range} shows the maximum detection range up to which ${\sim}100\%$ accuracy is achieved for various types of devices that we tested. 

\begin{table}[h!]
\centering
\caption{Performance analysis of threshold-based detection}
\begin{tabular}{||c | c ||} 
 \hline
 \textbf{Device type} & \textbf{Maximum range (m)} \\
 \hline
 IoT device (Arduino Uno) & 5 \\ 
 \hline
 Everyday electronics (desktop) & 3 \\
 \hline
 Cables (HDMI) & 22.5 \\
 \hline
\end{tabular}
\label{table_max_range}
\end{table}


\begin{table*}[h!]
\centering
\caption{Comparison of unintended RF emission detection methods}
\begin{tabular}{|| >{\centering}m{9em} | >{\centering}m{12em} | >{\centering}m{10em} | >{\centering}m{6em} | >{\centering}m{5em} | m{12em} ||} 
 \hline
 \textbf{Paper} & \textbf{Method} & \textbf{Target devices} & \textbf{Accuracy (\%)} & \textbf{Detection range (m)} & \textbf{Environment(s)}\\
 \hline
 Guardiola et al. \cite{dcomp1} & Nonparametric method & Two-way talk radios & 63.6 & 3 & Unspecified \\
 \hline
 Hegarty et al. \cite{dcomp2} & CNN & Arduino Uno & 100 & 0.3 & Lab \\
 \hline
 Vuagnoux et al. \cite{dcomp3} & Short Time Fourier Transform (STFT) & PS2 Keyboard & 95 & 20 & Anechoic chamber, office, \& residential building \\
\hline
Friedel et al. \cite{dcomp4} & Simple Additive Weighting (SAW) & IED & 86.1 & 0.2 & Anechoic chamber \\
\hline
Liu et al. \cite{dcomp5} & Correlation between external stimulus and spectral pattern change & Small camera (FLIR USB camera) & 93.23 & 0.1 & Lab \\
\hline
Alexander et al. \cite{dcomp6} & Emanation Model Analytics (SVM-based) & Arduino Uno, Raspberry Pi B+, Teensy LC & 100 & 0.3 & RF Shield Box \\
\hline
Mo et al. \cite{dcomp7} & Support Vector Machine (SVM) & LCD monitors & 98.95 & 0.1 & Semi-anechoic chamber \\
\hline
Hertenstein et al. \cite{dcomp8} & Cross-correlation and Hurst parameter thresholding & RC toy car, wireless doorbell & 97.6 (ROC) & 0.25 & Unspecified \\
\hline
Göksu et al. \cite{dcomp9} & Wavelet Packet Analysis (WPA) and MLP & Cars & 97 & 1 to 3 & Lab\\
\hline
Acharya et al. \cite{dcomp10} & Principal Component Analysis (PCA) & Two-way radios & Unspecified & 3 & Unspecified\\
\hline
Weng et al. \cite{dcomp11} & Neural Network or Multilayer Perceptron (MLP) & RC toy truck, wireless doorbell & 98.99 (Avg) & 10 & Anechoic chamber, office corridor\\
\hline
\textbf{This work} & \textbf{harmonic and IMP pattern-based detection} & \textbf{IoT devices, everyday electronics, and cables} & \textbf{100} & \textbf{22.5} & \textbf{Lab, office room, open space, \& anechoic chamber}\\
\hline
\end{tabular}
\label{table_sota_comparison}
\end{table*}

Applying our computational harmonic detector-based emanation detection method to our previously collected HDMI dataset (emanation from 3 types of HDMI cable in an office corridor \cite{date_hdmi}), ${\sim}100\%$ accuracy has been achieved up to \SI{22.5}{\meter} range. If we compare this result with our earlier CNN-based method, for iso-accuracy of ${\sim}100\%$, harmonic method has an extended detection range of \SI{22.5}{\meter}, compared to \SI{16}{\meter} for CNN approach. Also, for iso-distance of \SI{22.5}{\meter}, harmonic method has a better detection accuracy (${\sim}100\%$ vs ${\sim}95\%$). The following subsection compares our harmonic-based emanation detection method to other methods proposed in literature.  

\subsection{Performance Comparison}
Table~\ref{table_sota_comparison} compares our work with other notable works for unintended electromagnetic emission detection. Before delving into a detailed discussion, two essential factors must be considered regarding this comparison. Firstly, older devices used to have much higher EM emanation than newer ones. For example, PS2 keyboards (used in \cite{dcomp3}) had clock signals and often no shielding. So they used to have much higher emissions compared to modern USB keyboards that use differential signals (no clock) and shielding. Secondly, the primary goal of some of these works was not just to detect the emanation itself but to use it for other applications (covert communication \cite{dcomp2}, hidden camera detection \cite{dcomp5}, vehicle identification \cite{dcomp9}, etc.). Despite that, these works provide diverse insight towards emanation detection and are kept in this table.

The comparison is made in terms of 5 parameters: (1) detection method, (2) target devices, (3) accuracy, (4) maximum detection range, and (5) test environments. Several observations can be made from this table. Firstly, while other works focus on a specific type of device and a small test set, we have covered a wide range of devices (eight devices in three categories). Since emanation properties are hardware-dependent, a generalized, device-agnostic method is advantageous over device-specific methods in practical scenarios. Secondly, we have one of the best accuracies ($100\%$) reported among published works. While Hegarty et al. \cite{dcomp2} and Alexander et al. \cite{dcomp6} also achieved $100\%$ accuracy, their detection range was very short, well below a meter (\SI{0.3}{\meter}). Combining accuracy and maximum detection range, our method clearly renders better performance than others. Finally, our method is rigorously tested in different environments (anechoic chamber, lab, office room, open space out of office room, etc.) where RF background profiles vary widely. The only work that was found to perform this level of testing was done by Vuagnoux et al. \cite{dcomp3}.

\subsection{Device Agnostic Detection}
\label{subsec_device_agnostic}
The properties of emanation signals are hardware-dependent. So, an EM emanation detection method developed for one type of device will not work with other types. This limits the deployment feasibility of such methods. Our previous CNN-based method was also limited to the device data with which it was trained. However, the harmonic-based method is more device-agnostic. It has been tested to achieve $100\%$ accuracy for IoT devices (Arduino, PSoC, ESP32, and ZigBee), everyday electronics (desktop and monitor), and cables (HDMI and USB). Additionally, it has been tested for collected data from three other devices (RockPi 4 B+, webcam, and thumb drive) in our funding agency's testbed, where it also achieved ${\sim}100\%$ accuracy in emanation detection. So, it is of statistical significance to claim that our proposed harmonic-based emanation detection method is mostly device-agnostic.

\subsection{Emanation Detection in Other Environments}
\begin{figure*}
\centering
\includegraphics[width=7in]{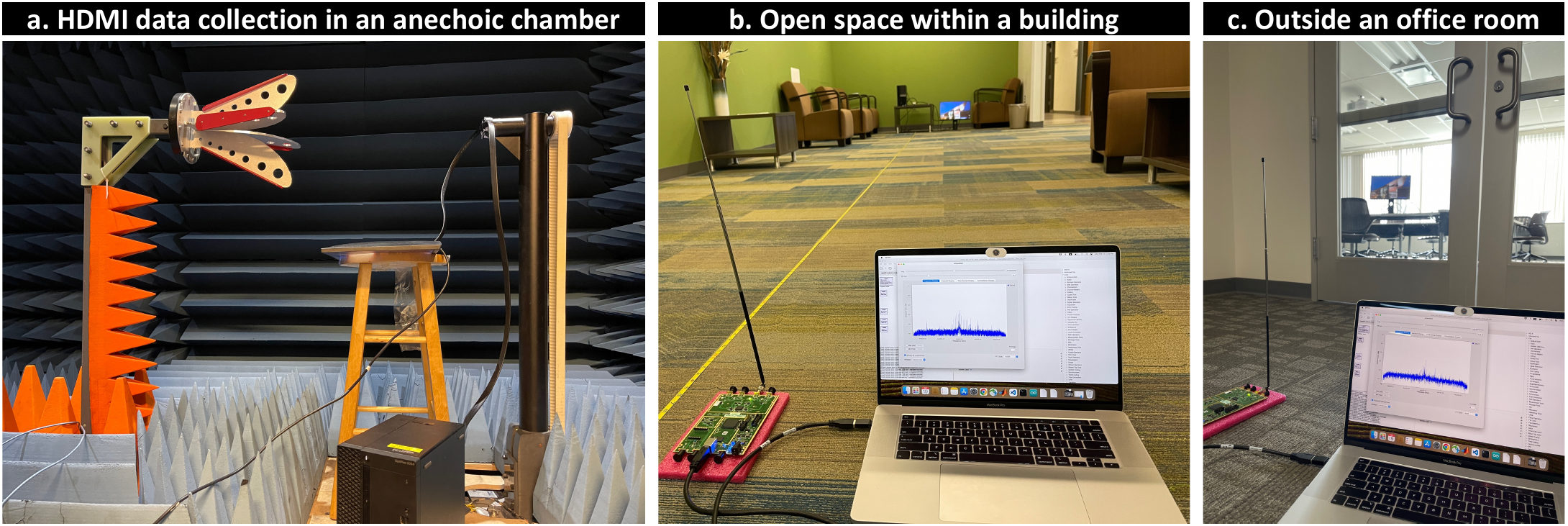}
\caption{To test the efficacy of the proposed algorithm in different environments, HDMI emanation data are being collected in (a) an anechoic chamber, (b) open space in a building (line-of-sight or LOS), and (c) outside of an office room (non-line-of-sight or NLOS).}
\label{fig_dev_agnostic_hdmi_env}
\end{figure*}

Environmental variation is a crucial factor as noise and interference levels change significantly. Also, EM emanation couples with long metallic elements to travel further. The presence of such elements also varies with the environment. To test our algorithm in different environments, an HDMI cable is used as the target in 3 new environments (in addition to the previous lab setup), which are shown in Fig.~\ref{fig_dev_agnostic_hdmi_env}.
\begin{itemize}
    \item \textbf{An anechoic chamber:} The whole setup (both target HDMI and receiver B210) is placed within the chamber. Data are collected using an extended USB cable connecting the B210 SDR from inside to a capturing Macbook outside. The maximum data collection range in our anechoic chamber is \SI{\sim5}{\meter}.
    \item \textbf{An open space within a building (line-of-sight transmission):} EM emanation can couple with various conducting elements surrounding the emanation source and travel further. An open space has limited coupling elements. Hence, the emanation range in such an environment is more conservative than others. However, due to the room size, the maximum range for data collection in this environment is \SI{11}{\meter}. 
    \item \textbf{Outside of an office room (non-line-of-sight transmission):} Here, the target HDMI (with the monitor and PC) is kept inside an office room. The receiving antenna, along with B210 SDR, was kept outside to mimic practical eavesdropping scenarios. The target setup is kept \SI{1.5}{\meter} away from the separating wall and door. Again, due to the building geometry, the outside data collection range was limited to \SI{6}{\meter}. Hence, the total separation between the target and receiver is \SI{7.5}{\meter}.
\end{itemize}

In all cases, our harmonic-based emanation detector renders ${\sim}100\%$ accuracy. This shows its robustness against environmental variability.

\section{Conclusion}
\label{conc}
In this work, we have extended our preliminary work on emanation detection by collecting data from a wide range of carefully chosen IoT devices (Arduino, PSoC, Zigbee, and ESP32), day-to-day electronic devices (PC and monitor), and cables (HDMI and USB). Data analysis reveals that each device’s emanation has a unique harmonic pattern with intermodulation products, unlike communication signals with fixed frequency bands, distinct spectra, and modulation patterns. Leveraging this, we propose a harmonic-based emanation detection method by developing a computational harmonic detector. The proposed method addresses all the weaknesses of the previously proposed CNN-based method. It can detect multiple emanation sources simultaneously, performs well even at low SNR (${\sim}1$ dB), has a lower computation cost, and is hardware-agnostic. This method provides ${\sim}100\%$ accuracy up to \SI{22.5}{\meter}, compared to ${\sim}95\%$ accuracy at the same distance using the CNN-based method. It also achieves ${\sim}100\%$ accuracy for all the tested devices, including IoT devices. The maximum detection ranges for such accuracy are reported. Furthermore, the performance of our method is compared with other emanation detection methods reported in the literature in terms of 5 parameters. Finally, the proposed method has been tested in 3 new environments to check its practical efficacy: in an anechoic chamber, an open space (direct line-of-sight transmission), and outside of an office room (non-line-of-sight transmission). This work paves the way for a smart and automated spectrum monitoring system to detect EM side-channel leakage from electronic devices.

\section*{Acknowledgments}
This research was supported by the Office of the Director of National Intelligence (ODNI), Intelligence Advanced Research Projects Activity (IARPA), via contract: 2021-21062400006. The views and conclusions contained herein are those of the authors and should not be interpreted as necessarily representing the official policies or endorsements, either expressed or implied, of the ODNI, IARPA, or the U.S. Government. The U.S. Government is authorized to reproduce and distribute reprints for Governmental purposes notwithstanding any copyright annotation thereon.




\vfill

\end{document}